\documentclass{article}
\usepackage[T1]{fontenc} 
\usepackage[utf8]{inputenc} 
\usepackage{ismir,amsmath,cite,url}
\usepackage{amsmath,amssymb}
\usepackage{graphicx}
\usepackage{color}
\usepackage{booktabs}
\usepackage{tabularx}
\usepackage[ruled,vlined]{algorithm2e}
\usepackage{float}

\usepackage{lineno}

\title{Music FaderNets: Controllable Music Generation Based On High-Level Features via Low-Level Feature Modelling}

\multauthor
{Hao Hao Tan$^1$ \hspace{1cm} Dorien Herremans$^1$} {
 $^1$ Singapore University of Technology and Design\\
{\tt\small \{haohao\_tan, dorien\_herremans\}@sutd.edu.sg}
}
\def\authorname{Hao Hao Tan, Dorien Herremans}

\begin{document}

\maketitle
\begin{abstract}
High-level musical qualities (such as emotion) are often abstract, subjective, and hard to quantify. Given these difficulties, it is not easy to learn good feature representations with supervised learning techniques, either because of the insufficiency of labels, or the subjectiveness (and hence large variance) in human-annotated labels. In this paper, we present a framework that can learn high-level feature representations with a limited amount of data, by first modelling their corresponding quantifiable \emph{low-level} attributes. We refer to our proposed framework as Music FaderNets, which is inspired by the fact that low-level attributes can be continuously manipulated by separate ``sliding faders'' through feature disentanglement and latent regularization techniques. High-level features are then inferred from the low-level representations through semi-supervised clustering using Gaussian Mixture Variational Autoencoders (GM-VAEs). Using arousal as an example of a high-level feature, we show that the ``faders'' of our model are disentangled and change linearly w.r.t. the modelled low-level attributes of the generated output music. Furthermore, we demonstrate that the model successfully learns the intrinsic relationship between arousal and its corresponding low-level attributes (rhythm and note density), with only \(1\%\) of the training set being labelled. Finally, using the learnt high-level feature representations, we explore the application of our framework in style transfer tasks across different arousal states. The effectiveness of this approach is verified through a subjective listening test. 
\end{abstract}
\section{Introduction}\label{sec:introduction}
We consider \textit{low-level} musical attributes as attributes that are relatively straightforward to quantify, extract and calculate from music, such as rhythm, pitch, harmony, etc. On the other hand, \textit{high-level} musical attributes refer to semantic descriptors or qualities of music that are relatively abstract, such as emotion, style, genre, etc. Due to the nature of abstractness and subjectivity in these high-level musical qualities, obtaining labels for these qualities typically requires human annotation. However, training conditional models on top of these human-annotated labels using supervised learning might result in sub-par performance because firstly, obtaining such labels can be costly, hence the amount of labels collected might be insufficient to train a model that can generalize well \cite{habib2019semi}; Secondly, the annotated labels could have high variance among raters due to the subjectivity of these musical qualities \cite{aljanaki2015emotion, ferreiralearning}.

    Instead of inferring high-level features directly from the music sample, we propose to use low-level features as a ``bridge'' between the music and the high level features. This is because the relationship between the sample and its low-level features can be learnt relatively easier, as the labels are easier to obtain. In addition, we learn the relationship between the low-level features and the high-level features in a data-driven manner. In this paper, we show that the latter works well even with a limited amount of labelled data. Our work relies heavily on the concept that each high-level feature is intrinsically related to a set of low-level attributes. By tweaking the levels of each low-level attribute in a constrained manner, we can achieve a desired change on the high-level feature. This idea is heavily exploited in rule-based systems \cite{bresin2000emotional, livingstone2010changing, ehrlich2019closed}, however rule-based systems are often not robust enough as their capabilities are constrained by the fixed set of predefined rules handcrafted by the authors. Hence, we propose an alternative path which is to \textit{learn} these implicit relationships with semi-supervised learning techniques.

To achieve the goals stated above, we intend to build a framework which can fulfill these two objectives:
\begin{itemize}
    \item Firstly, the model should be able to control multiple low-level attributes of the music sample in a continuous manner, as if it is controlled by sliding knobs on a console (or also known as \textit{faders}). Each knob should be independent from the others, and only controls one single feature that it is assigned to.
    
    \item Secondly, the model should be able to learn the relationship between the levels of the sliding knobs controlling the low-level features, and the selected high-level feature. This is analogous to learning a \textit{preset} of the sliding knobs on a console.
\end{itemize}

We named our model ``Music FaderNets'', with reference to musical ``faders'' and ``presets'' as described above. Achieving the first objective requires representation learning and feature disentanglement techniques. This motivates us to use \textit{latent variable models} \cite{kim2018tutorial} as we can learn separate latent spaces for each low-level feature to obtain disentangled controllability. Achieving the second objective requires the latent space to have a hierarchical structure, such that high-level information can be inferred from low-level representations. This is achieved by incorporating Gaussian Mixture VAEs \cite{jiang2016variational} in our model.
\section{Related Work}
\subsection{Controllable Music Generation}
The application of deep learning techniques for music generation has been rapidly advancing \cite{herremans2019emergence, briot2017deep, herremans2017functional, oore2018time, huang2018music}, however, embedding \textit{control} and \textit{interactivity} in these systems still remains a critical challenge \cite{briot2017deep}. Variants of conditional generative models (such as CGAN \cite{mirza2014conditional} and CVAE \cite{sohn2015learning}) are used to allow control during generation, which have attained much success mainly in the image domain. Fader Networks \cite{lample2017fader} is one of the main inspirations of this work (hence the name Music FaderNets), in which users can modify different visual features of an image using ``sliding faders''. However, their approach is built upon a CVAE with an additional adversarial component, which is very different from our approach. Recently, controllable music generation has gained much research interest, both on modelling low-level \cite{roberts2018hierarchical, hadjeres2017glsr, pati2019latent, engel2017latent} and high-level features \cite{dai2018music, choi2019encoding}. Specifically, \cite{hadjeres2017glsr} and \cite{pati2019latent} each proposed a novel latent regularization method to encode attributes along specific latent dimensions, which inspired the "sliding knob" application in this work.

\subsection{Disentangled Representation Learning for Music}
Disentangled representation learning has been widely used across both the visual \cite{chen2016infogan, higgins2017beta, kim2018disentangling, yingzhen2018disentangled} and speech domain \cite{hsu2017unsupervised, habib2019semi, wang2018style} to learn disjoint subsets of attributes. Such techniques have also been applied to music in several recent works, both in the audio \cite{luo2019learning, hung2018learning, hung2019musical} and symbolic domain \cite{brunner2018midi, yang2019deep, akamacontrolling}. The discriminator component in our model draws inspiration from both the explicit conditioning component in the EC\textsuperscript{2}-VAE model \cite{yang2019deep}, and the \textit{extraction} component in the Ext-Res model \cite{akamacontrolling}. We find that most of the work on disentanglement in symbolic music focuses on low-level features, and is done on monophonic music.

This research distinguishes itself from other related work through the following novel contributions:
\begin{itemize}
    \item We combine latent regularization techniques with disentangled representation learning to build a framework that can control various continuous low-level musical attribute values using ``faders'', and apply the framework on \emph{polyphonic} music modelling.
    \item We show that it is possible to infer high-level features from low-level latent feature representations, even under weakly supervised scenario. This opens up possibilities to learn good representations for abstract, high-level musical qualities even under data scarcity conditions. We further demonstrate that the learnt representations can be used for controllable generation based on high-level features.
    
\end{itemize}

\section{Proposed Framework}\label{sec:page_size}

\subsection{Gaussian Mixture Variational Autoencoders}
VAEs \cite{kingma2013auto} combine the power of both latent variable models and deep generative models, hence they provide both representation learning and generation capabilities. Given observations \(\textbf{X}\) and latent variables \(\textbf{z}\), the VAE learns a graphical model \(\textbf{z} \rightarrow \textbf{X}\) by maximizing the evidence lower bound (ELBO) of the marginal likelihood \(p(\textbf{X})\) as below:
\begin{equation*}
\mathcal{L}(p, q; \textbf{X}) = \mathbb{E}_{q(\textbf{z}|\textbf{X})}
[\textrm{log } p(\textbf{X}|\textbf{z})] - \mathcal{D}_{KL}(q(\textbf{z}|\textbf{X}) || p(\textbf{z}))
\end{equation*}
where \(q(\textbf{z}|\textbf{X})\) and \(p(\textbf{z})\) represent the learnt posterior and prior distribution respectively. In vanilla VAEs, \(p(\textbf{z})\) is an isotropic, unimodal Gaussian. Gaussian Mixture VAEs (GM-VAE) \cite{jiang2016variational} extend the prior to a mixture of \(K\) Gaussian components, which corresponds to learning a graphical model with an extra hierarchy of dependency \(c \rightarrow \textbf{z} \rightarrow \textbf{X}\). The newly introduced categorical variable \(c \in \mathcal{C}\), whereby \(|\mathcal{C}| = K\), is a discrete representation of the observations. Hence, a new distribution \(q(c|\textbf{X})\) is introduced to infer the class of each observation, which enables semi-supervised and unsupervised clustering applications.
 
Following \cite{jiang2016variational}, the ELBO of a GM-VAE is derived as:
\begin{equation*} \label{eq1}
\begin{split}
\mathcal{L}(p, q; \textbf{X}) &= \mathbb{E}_{q(\textbf{z}|\textbf{X})}[\textrm{log } p(\textbf{X}|\textbf{z})] \\
&- \displaystyle\sum_{k=1}^{K} q(c_k|\textbf{X})  \mathcal{D}_{KL}(q(\textbf{z}|\textbf{X}) || p(\textbf{z}| c_k)) \\
&- \mathcal{D}_{KL}(q(c|\textbf{X}) || p(c))
\end{split}
\end{equation*}

The original KL loss term from the vanilla VAE is modified into two new terms: (i) the KL divergence between the approximate posterior \(q(\textbf{z}|\textbf{X})\) and the conditional prior \(p(\textbf{z}|c_k)\), marginalized over all Gaussian components; (ii) the KL divergence between the cluster inferring distribution \(q(c|\textbf{X})\), and the categorical prior \(p(c)\).
\begin{figure*}[t!]
  \includegraphics[width=\textwidth, height=200pt, keepaspectratio]{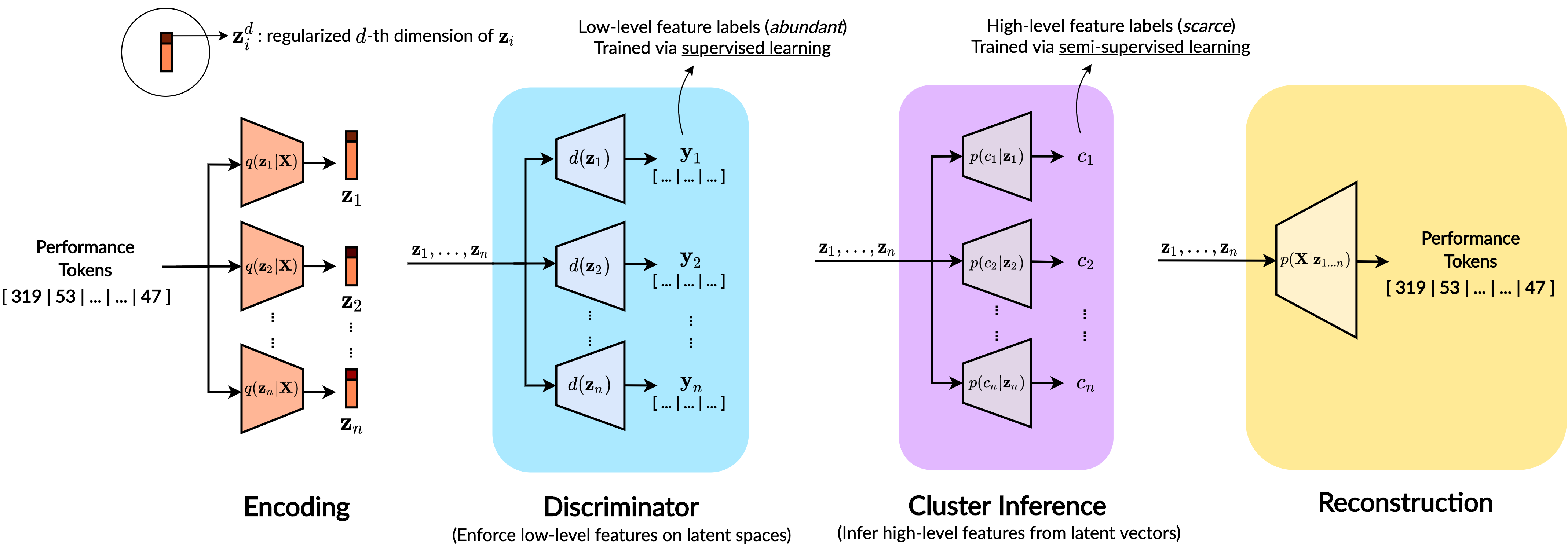}
  \caption{Music FaderNets model architecture.}
  \label{fig:architecture}
  \vspace{-12pt}
\end{figure*}

\subsection{Model Formulation}
Figure {\ref{fig:architecture}} shows the model formulation of our proposed Music FaderNets. Input \(\textbf{X}\) is a sequence of performance tokens converted from MIDI following \cite{oore2018time, huang2018music}. Assume that we want to model a high-level feature with \(K\) discrete states, which is related to a set of \(N\) low-level features. We denote the latent variables learnt for each low-level feature as \(\textbf{z}_{1...N}\); the labels for each low-level feature as \(\textbf{y}_{1...N}\); and the class inferred from each latent variable as \(c_{1...N}\).

The joint probability of \(\textbf{X}, \textbf{z}_{1...N}, c_{1...N}\) is written as:

\begin{equation*} \label{eq1}
\begin{split}
p(\textbf{X}, \textbf{z}_{1...N}, c_{1...N}) &= p(\textbf{X}|\textbf{z}_{1...N})\displaystyle \prod_{i=1}^{N} p(\textbf{z}_i|c_i)\displaystyle \prod_{i=1}^{N} p(c_i)
\end{split}
\end{equation*}

We assume that each categorical prior \(p(c_i)\), \(i \in [1, N]\) is uniformly distributed, and the conditional distributions \(p(\textbf{z}_i|c_i) = \mathcal{N}(\mu_{c_i}, \textrm{diag}(\sigma_{c_i}))\)
are diagonal-covariance Gaussians with learnable means and constant variances. For each low-level attribute, we learn an approximate posterior \(q(\textbf{z}_i |\textbf{X})\), parameteri\textbf{z}ed by an encoder neural network, that samples latent code \(\textbf{z}_i\) which represents the \(i\)-th low-level feature. 

The latent codes \(\textbf{z}_{1...N}\) are then passed through the remaining three components: (1) \textbf{Discriminator}: To ensure that \(\textbf{z}_i\) incorporates information of the assigned low-level feature, it is passed through a discriminator represented by a function \(d(\textbf{z}_i)\) to reconstruct the low-level feature label \(\textbf{y}_i\); (2) \textbf{Reconstruction}: All latent codes are fed into a global decoder network which parameterizes the conditional probability \(p(\textbf{X} | \textbf{z}_{1...n})\) to reconstruct the input \(\textbf{X}\); (3) \textbf{Cluster Inference}: This component parameterizes the cluster inference probability \(q(c|\textbf{X})\), with \(c\) representing the selected high-level feature. It can be approximated by \(q(c|\textbf{X}) \approx \mathbb{E}_{q(\textbf{z}|\textbf{X})}p(c|\textbf{z})\) \cite{hsu2018hierarchical}, where the cluster state is predicted from each latent code \(\textbf{z}_i\) instead of \(\textbf{X}\).

To incorporate the ``sliding knob'' concept, we need to map the change of value of an arbitrary dimension on \(\textbf{z}_i\) (denoted as \(\textbf{z}_i^d\), shown on Figure {\ref{fig:architecture}} as the darkened dimension) linearly to the change of value of the low-level feature label \(\textbf{y}_i\). After comparing across previous methods on conditioning and regularization \cite{hadjeres2017glsr, sohn2015learning, lample2017fader, pati2019latent}, we choose to adopt \cite{pati2019latent} which applies a latent regularization loss term written as \(\mathcal{L}_{\textrm{reg}}(\textbf{z}_i^d, \textbf{y}_i) = \textrm{MSE}(\textrm{tanh}(\mathcal{D}_{\textbf{z}_i^d}),\textrm{sign}(\mathcal{D}_{\textbf{y}_i}))\), where \(\mathcal{D}_{\textbf{z}_i^d}\) and \(\mathcal{D}_{\textbf{y}_i}\) denotes the \textit{distance matrix} of values \(\textbf{z}_i^d\) and \(\textbf{y}_i\) within a training batch respectively. We provide a detailed comparison study across each proposed method in Section \ref{sec:low-level}. Hence, if we define:
\vspace{-8pt}

\begin{equation} \label{eq1}
\mathcal{L}^{i}_{\phi}(p, q; \textbf{X})=\begin{cases}

\begin{split}
&\displaystyle\sum_{k=1}^{K} q(c_{i, k}|\textbf{X})  \mathcal{D}_{KL}(q(\textbf{z}_i|\textbf{X}) || p(\textbf{z}_i| c_{i,k}))\\
&+ \mathcal{D}_{KL}(q(c_i|\textbf{X}) || p(c_i)), \textrm{if unsupervised}
\end{split}\\ \\
 \mathcal{D}_{KL}(q(\textbf{z}_i|\textbf{X}) || p(\textbf{z}_i| c_i)), \textrm{if supervised}
\end{cases}
\end{equation}
then the entire training objective can be derived as:

\begin{equation} \label{eq1}
\begin{split}
\mathcal{L}(p, q; \textbf{X}) &= \mathbb{E}_{q(\textbf{z}_1|\textbf{X})...q(\textbf{z}_N|\textbf{X})}[\textrm{log } p(\textbf{X}|\textbf{z}_1, \textbf{z}_2, ..., \textbf{z}_N)] \\
&- \beta \cdot \displaystyle\sum_{i=1}^{N} \mathcal{L}^{i}_{\phi}(p, q; \textbf{X}) + \displaystyle\sum_{i=1}^{N} \mathcal{L}_{\textrm{reg}}(\textbf{z}_i^d, \textbf{y}_i)
\\
&+ \mathbb{E}_{q(\textbf{z}_1|\textbf{X})...q(\textbf{z}_n|\textbf{X})}[\textrm{log } p(\textbf{y}_1|\textbf{z}_1)...p(\textbf{y}_N|\textbf{z}_N)] \\
\end{split}
\end{equation}
where \(\beta\) is the KL weight hyperparameter \cite{higgins2017beta}. The first term in Eq. \ref{eq1} represents the reconstruction loss. The second KL loss term (derived from the ELBO function of GM-VAE) correspond to the cluster inference component, which allows both \textit{supervised} and \textit{unsupervised} training setting, depending on the availability of label \(c\). If we omit the cluster inference component, it could conform to a vanilla VAE by replacing this term with the KL loss term of VAE.  The third term is the latent regularization loss applied during the encoding process. The last term is the reconstruction loss of the low-level feature labels, which corresponds to the discriminator component. All encoders and decoders are implemented with gated recurrent units (GRUs), and teacher-forcing is used to train all decoders.

\section{Experimental Setup}\label{sec:typeset_text}
In this work, we chose \textit{arousal} (which refers to the energy level conveyed by the song \cite{russell1980circumplex}) as the high-level feature to be modelled. In order to select relevant low-level features, we refer to musicology papers such as \cite{ehrlich2019closed, gabrielsson2001influence, gomez2007relationships}, which suggest that arousal is related to features including rhythm density, note density, key, dynamic, tempo, etc. Among these low-level features, we focus on modelling the score-level features in this work (i.e. rhythm, note and key). 

\subsection{Data Representation and Hyperparameters}

We use two polyphonic piano music datasets for training: the \textbf{Yamaha Piano-e-Competition dataset} \cite{oore2018time}, and the \textbf{VGMIDI dataset}\cite{ferreiralearning}, which contains piano arrangements of 95 video game soundtracks in MIDI, annotated with valence and arousal values in the range of -1 to 1. The arousal labels are used to guide the cluster inference component in our GM-VAE model using semi-supervised learning. We extract every 4-beat segment from each music sample, with a beat resolution of 4 (quarter-note granularity). Each segment is encoded into event-based tokens following \cite{oore2018time} with a maximum sequence length of 100. This results in a total of 103,934 and 1,013 sequences from the Piano e-Competition and VGMIDI dataset respectively, which are split into train/validation/test sets with a ratio of 80/10/10.

Inspired by \cite{yang2019deep}, we represent each rhythm label, \(\textbf{y}_{\textrm{rhythm}}\), as a sequence of 16 one-hot vectors with 3 dimensions, denoting an onset for any pitch, a holding state, or a rest. The \textit{rhythm density} value is calculated as the number of onsets in the sequence divided by the total sequence length. Each note label, \(\textbf{y}_{\textrm{note}}\), is represented by a sequence of 16 one-hot vectors with 16 dimensions, each dimension denoting the number of notes being played or held at that time step (we assume a minimum polyphony of 0 and a maximum of 15). The \textit{note density} value is the average number of notes being played or held for per time step. For key, we use the key analysis tool from music21 \cite{cuthbert2010music21} to extract the estimated global key of each 4-beat segment. The key is represented using a 24-dimension one-hot vector, accounting for major and minor modes. In this work, we directly concatenate the key vector as a conditioning signal with \(\textbf{z}_{\textrm{rhythm}}\)  and \(\textbf{z}_{\textrm{note}}\) as an input to the global decoder for reconstruction. For representing arousal, we split the arousal ratings into two clusters \((K=2)\): \textit{high} arousal cluster for positive labels, and \textit{low} arousal cluster for negative labels. We remove labels annotated within the range [-0.1, 0.1] so as to reduce ambiguity in the annotations.

The hyperparameters are tuned according to the results on the validation set using grid search. The mean vectors of \(p(c|\textbf{z})\) are all randomly initialized with Xavier initialization, whereas the variance vectors are kept fixed with value \(e^{-  2}\). We observe that the following annealing strategy for \(\beta\) leads to the best balance between reconstruction and controllability: \(\beta\) is set to 0 in the first 1,000 training steps, and is slowly annealed up to 0.2 in the next 10,000 training steps. We set the batch size to 128, all hidden sizes to 512, and the encoded \(\textbf{z}\) dimensions to 128. The Adam optimizer is used with a learning rate of \(10^{-3}\).

\subsection{Measuring the Controllability of Latent Features}\label{sec:low-level}

\begin{table*}[t!]
\small
\resizebox{\textwidth}{!}{
\begin{tabular}{@{}ccccccc@{}}
\toprule
  & \multicolumn{2}{c}{Consistency} & \multicolumn{2}{c}{Restrictiveness} & \multicolumn{2}{c}{Linearity}   \\ \midrule
  & Rhythm Density & Note Density   & Rhythm Density   & Note Density     & Rhythm Density & Note Density   \\ \midrule
Proposed (Vanilla VAE) & 0.4367 $\pm$ 0.0258 & 0.3490 $\pm$ 0.0360 & 0.6645 $\pm$ 0.0169   & 0.6481 $\pm$ 0.0154   & \textbf{0.7805 $\pm$ 0.0142} & \textbf{0.8255 $\pm$ 0.0107} \\
Proposed (GM-VAE) & \textbf{0.5096 $\pm$ 0.0248} & 0.4207 $\pm$ 0.0309 & 0.6603 $\pm$ 0.0164   & 0.6457 $\pm$ 0.0132   & 0.7580 $\pm$ 0.0124 & 0.7792 $\pm$ 0.0177 \\ 
Pati et al.  \cite{pati2019latent}     & 0.4625 $\pm$ 0.0264 & \textbf{0.5100 $\pm$ 0.0150} & 0.6417 $\pm$ 0.0171   & 0.5497 $\pm$ 0.0206   & 0.7613 $\pm$ 0.0171 & 0.8220 $\pm$ 0.0143 \\
CVAE \cite{sohn2015learning}             & 0.2613 $\pm$ 0.0376 & 0.4997 $\pm$ 0.0355 & \textbf{0.6863 $\pm$ 0.0221}   & 0.7140 $\pm$ 0.0130   & 0.4969 $\pm$ 0.0166 & 0.3997 $\pm$ 0.0411 \\
Fader Networks \cite{lample2017fader}    & 0.2730 $\pm$ 0.0366 & 0.4983 $\pm$ 0.0425 & 0.6861 $\pm$ 0.0163   & \textbf{0.7379 $\pm$ 0.0149}   & 0.5482 $\pm$ 0.0283 & 0.4647 $\pm$ 0.0292 \\
GLSR   \cite{hadjeres2017glsr}            & 0.1891 $\pm$ 0.0346 & 0.1969 $\pm$ 0.0831 & 0.6365 $\pm$ 0.0276   & 0.7136 $\pm$ 0.0185  & 0.2465 $\pm$ 0.0197 & 0.1799 $\pm$ 0.0209 \\
\bottomrule
\end{tabular}
}
\label{tab:low-level}
\vspace{-12pt}
\caption{Experimental results (conducted on the Yamaha dataset test split) on the controllability of low-level features (rhythm density and note density) using disentangled latent variables. Bold marks the best performing model. }
\end{table*}

The proposed Music FaderNets model should meet two requirements: (i) Each ``fader'' independently controls one low-level musical feature without affecting other features (disentanglement), and (ii) the ``faders'' should change linearly with the controlled attribute of the generated output (linearity). For disentanglement, we follow the definition proposed in 
\cite{shu2019weakly} which decomposes the concept of disentanglement into \textit{generator consistency} and \textit{generator restrictiveness}. Using rhythm density as an example:
\begin{itemize}
    \item \emph{Consistency} on rhythm density means that for the same value of \(\textbf{z}_{\textrm{rhythm}}^{d}\), the value of the output's rhythm density should be consistent.
    \item \emph{Restrictiveness} on rhythm density means that changing the value of \(\textbf{z}_{\textrm{rhythm}}^{d}\) does not affect the attributes other than rhythm density (in our case, note density).
    \item \emph{Linearity} on rhythm density means that the value of rhythm density is directly proportional to the value of \(\textbf{z}_{\textrm{rhythm}}^{d}\), which is analogous to a sliding fader. 
\end{itemize}

We will be evaluating all three of these points in our experiment. 
For evaluating linearity, \cite{pati2019latent} proposed a slightly modified version of the interpretability metric by \cite{adel2018discovering}, which includes the following steps: (1) encode each sample in the test set, obtain the rhythm latent code and the dimension \(\textbf{z}^d\) which has the maximum mutual information with regards to the attribute; (2) learn a linear regressor to predict the input attribute values based on \(\textbf{z}^d\). The linearity score is hence the coefficient of determination (\(R^2\)) score of the linear regressor. However, this method evaluates only the encoder and not the decoder. As we want the sliding knobs to directly impact the output, we argue that the relationship between \(\textbf{z}^d\) and the output attributes should be more important. Hence, we propose to ``slide'' the values of the regularized dimension \(\textbf{z}^d\) within a given range and decode them into reconstructed outputs. Then, instead of predicting the \textit{input} attributes given the encoded \(\textbf{z}^d\), the linear regressor learns to predict the corresponding \textit{output} attributes given the ``slid'' values of \(\textbf{z}^d\).

\begin{figure}
  \includegraphics[width=\columnwidth]{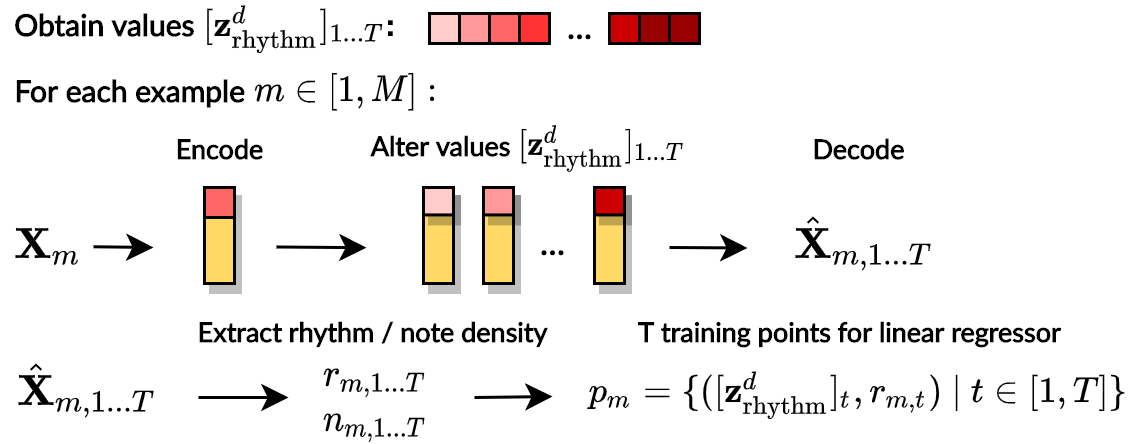}
  \vspace{-20pt}
  \caption{Workflow of obtaining evaluation metrics for ``faders'' controlling rhythm density.}
  \label{fig:evaluation}
  \vspace{-10pt}
\end{figure}

We demonstrate a single workflow to calculate the consistency, restrictiveness and linearity scores of a given model based on the low-level features (we use rhythm density as an example low-level feature for the discussion below), as depicted in Figure \ref{fig:evaluation}. After obtaining the rhythm density latent code for all samples in the training set and finding the minimum and maximum value of \(\textbf{z}_{\textrm{rhythm}}^d\), we ``slide'' for \(T=8\) steps by calculating \(\min(\textbf{z}_{\textrm{rhythm}}^d) + \frac{t}{T}(\max(\textbf{z}_{\textrm{rhythm}}^d) - \min(\textbf{z}_{\textrm{rhythm}}^d)), \textrm{with } t \in [1, T]\). This results in a list of values denoted as \([\textbf{z}_{\textrm{rhythm}}^d]_{1...T}\). Then, we conduct the following steps:
\begin{enumerate}\setlength{\itemsep}{0pt} \setlength{\parskip}{0pt}
    \item Randomly select \(M=100\) samples from the test set, and encode each sample into \(\textbf{z}_\textrm{rhythm}\) and \(\textbf{z}_\textrm{note}\);
    \item Alter the \(d\)-th element in \(\textbf{z}_\textrm{rhythm}\) using the values in the range \([\textbf{z}_{\textrm{rhythm}}^d]_{1...T}\), to obtain \([\hat{\textbf{z}}_\textrm{rhythm}]_{m, 1...T}\) for each sample $m$;
    \item Decode each new rhythm density latent code together with the unchanged note density latent code \(\textbf{z}_\textrm{note}\) to get \(\hat{\textbf{X}}_{m, 1...T}\);
    \item Calculate rhythm density \(r_{m, 1...T}\) and note density \(n_{m, 1...T}\) for each reconstructed output;
    \item Pair up the new rhythm density latent code with the resulting rhythm density of the output as \(T\) training data points \(p_m = \{([\textbf{z}_{\textrm{rhythm}}^d]_{t}, r_{m, t}) \ | \ t \in [1, T]\}\) for a linear regressor. 
\end{enumerate}
The final evaluation scores are then calculated as follows: 
\begin{equation}
\label{eq:scores1}
\textrm{Consistency score} = 1 - \frac{1}{T} \displaystyle\sum_{t=1}^{T} \underset{t}{\sigma} (r_{1...M, t})
\end{equation}
\begin{equation}
\label{eq:scores2}
\textrm{Restrictiveness score} = 1 - \frac{1}{M} \displaystyle\sum_{m=1}^{M} \underset{m}{\sigma} (n_{m, 1..T})
\end{equation}
\begin{equation}
\label{eq:scores3}
\textrm{Linearity score} = R^2(\mathcal{M}(p_{1...M}))
\end{equation}
where \(\sigma(\cdot)\) denotes the standard deviation, and \(\mathcal{M}\) denotes the linear regressor model. In other words, consistency calculates the average standard deviation across all output rhythm density values given the same \(\textbf{z}_{\textrm{rhythm}}^d\), whereas restrictiveness calculates the average standard deviation across all output note density values given the changing \(\textbf{z}_{\textrm{rhythm}}^d\). In a perfectly disentangled and linear model, the consistency, restrictiveness and linearity scores should be equal to 1, and higher scores indicate better performance.

\section{Experiments and Results}

We compare the evaluation scores of our proposed model, using both a vanilla VAE (omitting the cluster inference component) and GM-VAE, with several models proposed in related work on controllable synthesis: CVAE \cite{sohn2015learning},  Fader Networks \cite{lample2017fader}, GLSR \cite{hadjeres2017glsr} and Pati et al. \cite{pati2019latent}. We repeat the above steps for 10 runs for each model and report the mean and standard deviation of each score. Table 1 shows the evaluation results. Overall, our proposed models achieve a good all-rounded performance on every metric as compared to other models, especially in terms of linearity, models that use \cite{pati2019latent}'s regularization method largely outperform other models. Our model shares similar results with \cite{pati2019latent}, however as compared to their work, we encode a multi-dimensional, regularized latent space instead of a single dimension value for each low-level feature, thus allowing more flexibility. Our model can also be used for ``generation via analogy'' as mentioned in EC\textsuperscript{2}-VAE \cite{yang2019deep}, by mix-matching \(\textbf{z}_\textrm{rhythm}\) from one sample with \(\textbf{z}_\textrm{note}\) from another. Moreover, the feature latent vectors can be used to infer interpretable and semantically meaningful clusters.

\subsection{Inferring High-Level Features from Latent Low-Level Representations}\label{sec:infer-latent}

\begin{figure}
  \includegraphics[width=\columnwidth]{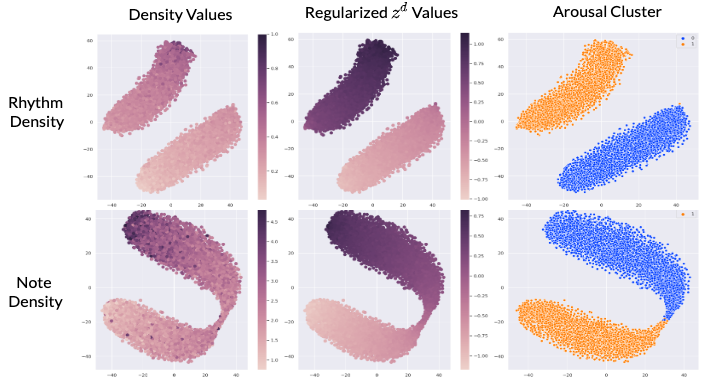}
  \vspace{-10pt}
  \caption{Visualization of rhythm (top) and note (bottom) density latent space in the GM-VAE. Each column is colored in terms of: (left) original density values, (middle) regularized \(\textbf{z}^d\) values, (right) arousal cluster labels (0 refers to low arousal and 1 refers to high arousal).}
  \label{fig:high-level}
  \vspace{-10pt}
\end{figure}

Figure \ref{fig:high-level} visualizes the rhythm and note density latent space learnt by GM-VAE using t-SNE dimensionality reduction. We observe that both spaces successfully learn a Gaussian-mixture space with two well-separated components, which correspond to high and low arousal clusters, even though it was trained with only around 1\% of labelled data. We also find that the regularized \(\textbf{z}^d\) values capture the overall trend of the actual rhythm and note density values. Interestingly, the model learns the implicit relationship between high/low arousal and the corresponding levels of rhythm/note density. From Figure \ref{fig:high-level}, we observe that the high arousal cluster corresponds to higher rhythm density and lower note density, whereas the low arousal cluster corresponds to lower rhythm density and higher note density. This is reasonable as music segments with high arousal often consist of fast running notes and arpeggios, being played one note at a time, whereas music segments with low arousal often exhibit a chordal texture with more sustaining notes and relatively less melodic activity. 

\begin{figure}
  \includegraphics[width=\columnwidth]{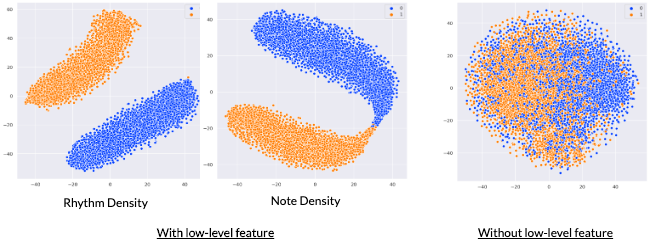}
  \vspace{-10pt}
  \caption{Arousal cluster visualization of GM-VAE with (left), and without (right) using low-level features.}
  \label{fig:cluster-low-high}
  \vspace{-8pt}
\end{figure}

To further inspect the importance of using low-level features, we train a separate GM-VAE model with only one encoder (without discriminator component), which encodes only a single latent vector for each segment. The model is trained to infer the arousal label with the single latent vector similarly in a semi-supervised manner, and the hyperparameters are kept the same. From Figure \ref{fig:cluster-low-high}, we can observe that the latent space learnt without using low-level features is not well-segregated into two separate components, suggesting that the right choice of low-level features helps the learning of a more discriminative and disentangled feature latent space.

The major advantage demonstrated from the results above is that by carefully choosing low-level features supported by domain knowledge, semi-supervised (or weakly supervised) training can be leveraged to learn interpretable representations that can capture implicit relationships between high-level and low-level features, overcoming the difficulties mentioned in the introduction section. This is an important insight for learning representations of abstract musical qualities under label scarcity conditions in future.

\begin{figure}
  \includegraphics[width=\columnwidth]{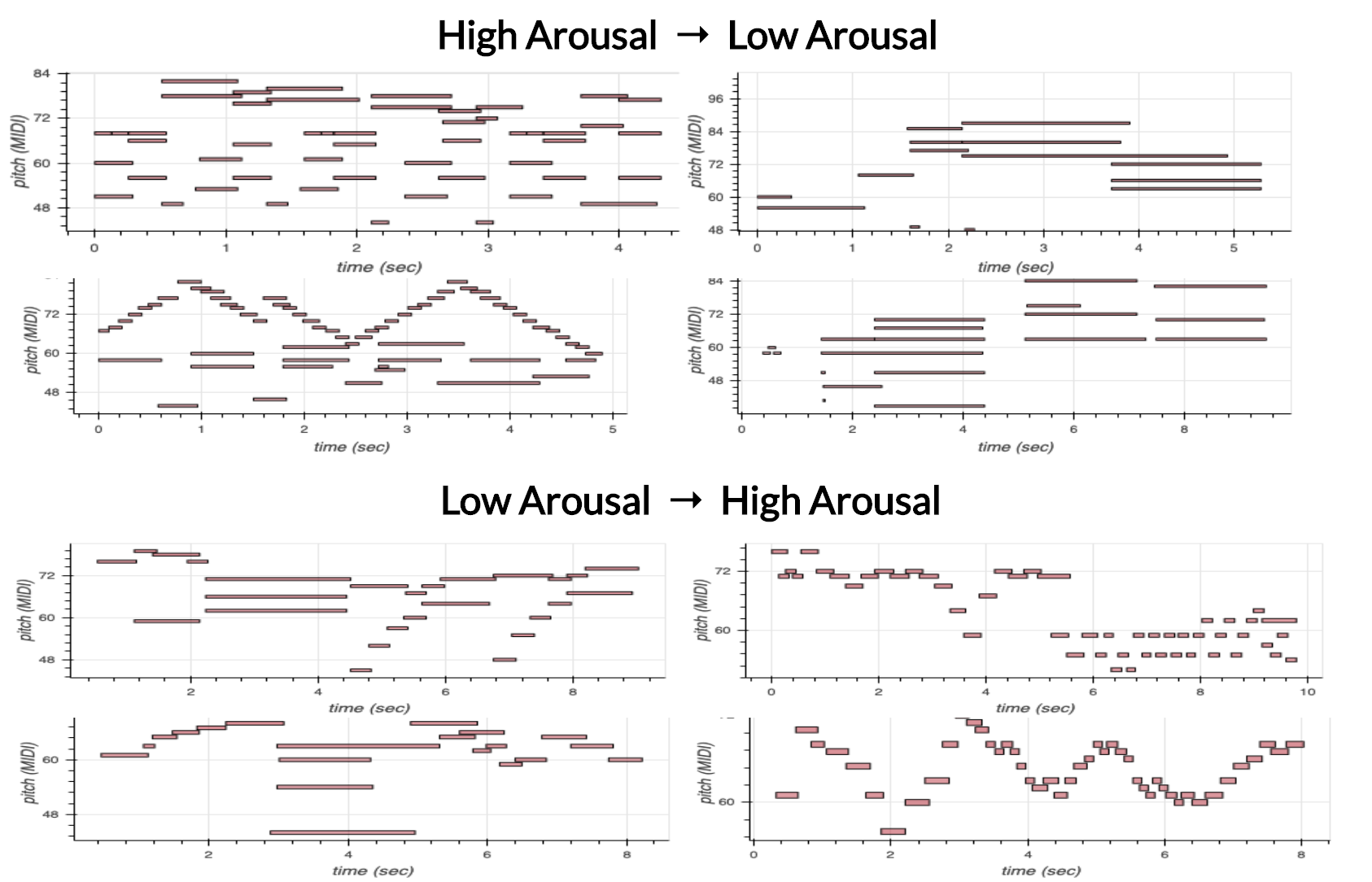}
  \caption{Examples of arousal transfer on music samples.}
  \label{fig:arousal-transfer}
  \vspace{-10pt}
\end{figure}

\begin{figure}[t]
  \includegraphics[width=\columnwidth]{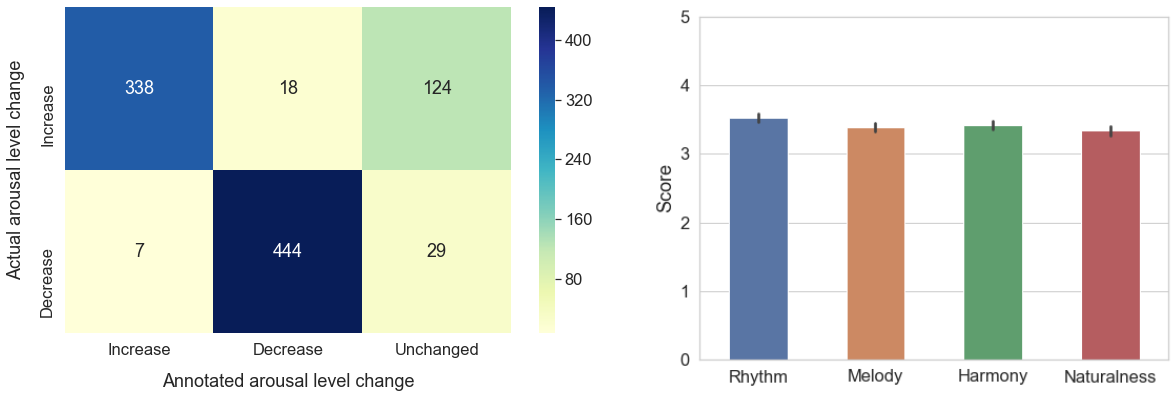}
  \vspace{-10pt}
  \caption{Subjective listening test results. Left: Heat map of annotated arousal level change against actual arousal level change. Right: Bar plot of opinion scores for each musical quality, with 95\% confidence interval.}
  \vspace{-6pt}
  \label{fig:subjective-results}
\end{figure}

\subsection{Style Transfer on High Level Features}
Utilizing the learnt high-level feature representations enables the application of feature style transfer. Following \cite{luo2019learning}, given the means of each Gaussian component, \(\mu_\textrm{arousal=0}\) and \(\mu_\textrm{arousal=1}\), the ``shifting vector'' from high arousal to low arousal is \(s_\textrm{low\_shift} = \mu_\textrm{arousal=0} - \mu_\textrm{arousal=1}\), and vice versa. To shift a music segment from high to low arousal, we modify the latent codes by \(\textbf{z}^{\prime}_\textrm{rhythm} = \textbf{z}_\textrm{rhythm} + s_\textrm{low\_shift}, \ \textbf{z}^{\prime}_\textrm{note} = \textbf{z}_\textrm{note} + s_\textrm{low\_shift}\). Both new latent codes \(\textbf{z}^{\prime}_\textrm{rhythm}\) and \(\textbf{z}^{\prime}_\textrm{note}\) are fed into the global decoder for reconstruction. For cases where \(c_\textrm{rhythm} \neq c_\textrm{note}\), we choose to perform shifting only on the latent codes which are not lying within the target arousal cluster. Figure \ref{fig:arousal-transfer} shows several examples of arousal shift performed on given music segments. We can observe that the shift is clearly accompanied with the desired changes in rhythm density and note density, as mentioned in Section \ref{sec:infer-latent}. More examples are available online.\footnote{\url{https://music-fadernets.github.io/}} We also conducted a subjective listening test to evaluate the quality of arousal shift performed by Music FaderNets. We randomly chose 20 music segments from our dataset, and performed a low-to-high arousal shift on 10 segments and a high-to-low arousal shift on the other 10. Each subject listened to the original sample and then the transformed sample, and was asked whether (1) the arousal level changes after the transformation, and; (2) how well the transformed sample sounds in terms of rhythm, melody, harmony and naturalness, on a Likert scale of 1 to 5 each.

A total of 48 subjects participated in the survey. We found that 81.45\% of the responses agreed with the actual direction of level change in arousal, shifted by the model. This showed that our model is capable of shifting the arousal level of a piece to a desired state. From the heat map shown in Figure \ref{fig:subjective-results}, we observe that shifting from high to low arousal has a higher rate of agreement (92.5\%) than shifting from low to high arousal (70.41\%). Meanwhile, the mean opinion score of rhythm, melody, harmony and naturalness were reported at 3.53, 3.39, 3.41 and 3.33 respectively, showing that the quality of the generated samples are generally above moderate level.

\section{Conclusion and Future Work}

We propose a novel framework called Music FaderNets\footnote{Source code available at: \url{https://github.com/gudgud96/music-fader-nets}}, which can generate new variations of music samples by controlling levels (``sliding knobs'') of low-level attributes, trained with latent regularization and feature disentanglement techniques. We also show that the framework is capable of inferring high-level feature representations (``presets'', e.g. arousal) on top of latent low-level representations by utilizing the GM-VAE framework. Finally, we demonstrate the application of using learnt high-level feature representations to perform arousal transfer, which was confirmed in a user experiment. The key advantage of this framework is that it can learn interpretable mixture components that reveal the intrinsic relationship between low-level and high-level features using semi-supervised learning, so that abstract musical qualities can be quantified in a more concrete manner with limited amount of labels. 

While the strength of arousal transfer is gradually increased, we find that the identity of the original piece is also gradually shifted. A recent work on text generation using VAEs \cite{xu2019variational} observed this similar trait and attributed its cause to the ``latent vacancy" problem by topological analysis. A possible solution is to adopt the Constrained-Posterior VAE \cite{xu2019variational}, in which we aim to explore in future work. Future work will also focus on applying the framework on other sets of abstract musical qualities (such as valence \cite{russell1980circumplex}, tension \cite{herremans2017morpheus}, etc.), and extending the framework to model multi-track music with longer duration to produce more complete music.

\newpage

\section{Acknowledgements}

We would like to thank the anonymous reviewers for their
constructive reviews. We also thank Yin-Jyun Luo for the insightful discussions on GM-VAEs. This work is supported by MOE Tier 2 grant no. MOE2018-T2-2-161 and SRG ISTD 2017 129. The subjective listening test is approved by the Institutional Review Board under SUTD-IRB 20-315. We would also like to thank the volunteers for taking the subjective listening test.

\bibliography{ISMIRtemplate}

%
%
%
%

\end{document}